\documentstyle[12pt,epsf]{article}
\begin{document}
\baselineskip 16pt plus 2pt minus 2pt
\newcommand{\beq}{\begin{equation}}
\newcommand{\eeq}{\end{equation}}
\newcommand{\beqa}{\begin{eqnarray}}
\newcommand{\eeqa}{\end{eqnarray}}
\renewcommand{\thefootnote}{\#\arabic{footnote}}
\newcommand{\ve}{\varepsilon}
\newcommand{\eps}{\epsilon}
\newcommand{\krig}[1]{\stackrel{\circ}{#1}}
\newcommand{\barr}[1]{\not\mathrel #1}

\begin{titlepage}

\hfill {\bf  \today}

%\noindent SUBMITTED VERSION, \today
\hfill TK 95 24

\hfill MKPH-T-95-25

\hfill hep--ph/9509nnn

\vspace{1.0cm}

\begin{center}

{\large { \bf THE STRANGENESS RADIUS AND MAGNETIC MOMENT OF THE
    NUCLEON REVISITED \footnote{Work supported in part by the
    Deutsche Forschungsgemeinschaft (SFB 201)}}}
%UPDATE ON ISOSCALAR
%STRANGE MATRIX--ELEMENTS IN THE NUCLEON

\vspace{1.2cm}

H.-W. Hammer$^{\ddag}$\footnote{electronic address:
hammerhw@vkpmzp.kph.uni-mainz.de},
Ulf-G. Mei\ss ner$^{\dag}$\footnote{electronic address:
meissner@pythia.itkp.uni-bonn.de},
D. Drechsel$^{\ddag}$\footnote{electronic address:
drechsel@vkpmzp.kph.uni-mainz.de}

\vspace{0.8cm}

$^{\ddag}$Universit\"at Mainz, Institut f\"ur Kernphysik, J.-J.-Becher
Weg 45\\ D--55099 Mainz, Germany

\vspace{0.3cm}
$^{\dag}$Universit\"at Bonn, Institut f\"ur Theoretische Kernphysik, Nussallee
14-16,\\ D--53115 Bonn, Germany

\vspace{0.4cm}

\end{center}

\vspace{1.5cm}

\begin{abstract}
\noindent We update Jaffe's estimate of the strange isoscalar radius
and magnetic moment of the nucleon. We make use of a recent
dispersion--theoretical fit to the nucleon electromagnetic form
factors and an improved description of symmetry breaking in the vector
nonet. We find $\mu_s = -0.24 \pm 0.03$~n.m. and
$r_s^2 = 0.21 \pm 0.03$~fm$^2$. The strange formfactor $F_2^s (t)$
follows a dipole with a cut--off mass of 1.46~GeV,
$F_2^s (t)= \mu_s (1-t/2.14 \, {\rm GeV}^2 )^{-2}$.
These numbers should be considered as upper limits on the strange
vector current matrix--elements in the nucleon.
\end{abstract}

\vspace{2cm}

\vfill

\end{titlepage}

%\noindent {\bf 1.}
\section{Introduction}
Over the last years, there has been considerable activity to pin down
strange matrix--elements in the nucleon. Much interest has been focused
on the strange quark content to the proton spin
as measured in DILS and the strange quark
contribution to the nucleon mass as revealed in the analysis of the
pion--nucleon $\sigma$--term. Jaffe \cite{bob} estimated the matrix
elements of the operators $r_s^2 \equiv s^\dagger (\vec x)
{\vec x}^2 s(\vec x)$ and
$\mu_s \equiv \vec{x} \times \bar{s} \vec{\gamma} s$
using the dispersion theory fits to the nucleon isoscalar form factors
of H\"ohler et al. \cite{hoeh76}. Such signals of strangeness in the
nucleon have also been considered in a variety of hadron models and
have lead to dedicated experiments like SAMPLE at MIT-Bates, a flurry
of CEBAF proposals, one proposal at MAMI and many other
experimental as well as
theoretical activities (for a review, see Ref.\cite{musolf}).
In this letter, we want to update the estimate of the strange
magnetic moment and radius by incorporating various new
developments not available at the time Ref.\cite{bob} was
written. First, a new dispersion theoretical analysis of the nucleon
form factors has been performed \cite{MMD}.
It improves upon the work of H\"ohler
et al. \cite{hoeh76} in various respects. These are the implementation
of the constraints from perturbative QCD (pQCD) at large momentum transfer,
the inclusion of the recent neutron--atom scattering length
determination \cite{kopek} to constrain the neutron charge radius
and, of course, the inclusion of new data at low, moderate and high
momentum transfer (as listed in \cite{MMD}). In that paper strong support
for the basic assumption of Jaffe's analysis, namely the
identification of the second pole in the isoscalar Dirac and Pauli formfactors
with the $\phi (1020)$, is presented. In fact, even the
location of the third pole
necessary in the isoscalar channel could be identified with the mass
of the $\omega (1600)$ (denoted by $S'$ in \cite{MMD}). Furthermore,
the symmetry breaking of the strong and electroweak interactions
in the vector nonet has been considerably refined \cite{joe} leading to
an improved value of the $\omega \phi$ mixing angle $\epsilon$.
These are the ingredients we will use to update and sharpen the analysis of
Jaffe.

%\bigskip
\section{Formalism}
%\noindent {\bf 2.}
It is straightforward to generalize Jaffe's
parametrization of the isoscalar formfactors to account for the
constraints from pQCD. We follow Ref.\cite{MMD} and separate the
spectral functions of the pertinent form factors into a hadronic
(meson pole) and a quark (pQCD) component as follows,
\beq
F_i^{(I=0)} (t) = \tilde{F}_i^{(I=0)} (t) L(t) = \Bigg[ \sum_{I=0}
\frac{a_I^{(I=0)} \, L^{-1}(M^2_{(I=0)})}{M^2_{(I=0)} - t }\Bigg] \, \Bigg[
\ln \Bigg( \frac{\Lambda^2 - t}{Q_0^2} \Bigg)\Bigg]^{-\gamma}
\label{ffism} \eeq
with
\begin{equation}
L(t) = \biggl[ \ln \bigg(\frac{\Lambda^2 - t}{Q_0^2} \biggr)
\biggr]^{-\gamma} \, \, \, .
\label{Lt}
\end{equation}
Here, $\Lambda \simeq 10$ GeV$^2$ separates the hadronic from the
quark contributions, $Q_0$ is related to $\Lambda_{\rm QCD}$
and $\gamma$ is the anomalous dimension,
\begin{equation}
F_i (t) \to (-t)^{-(i+1)} \, \biggl[ \ln\biggl(\frac{-t}{Q_0^2}\biggr)
\biggr]^{-\gamma} \, , \quad \gamma = 2 + \frac{4}{3\beta}
\, \, , \quad i = 1,2 \, \, ,
\label{fasy}
\end{equation}
with $\beta$ the QCD $\beta$--function and $t$ the invariant momentum
transfer squared. For the best fit to the available proton {\it and}
neutron data, the isoscalar masses are
$M_\omega=0.782\mbox{ GeV}$, $M_\phi=1.019\mbox{ GeV}$ and
$M_{S'}=1.60\mbox{ GeV}$
with the corresponding residua being $a_1^\omega = 0.747$,
$a_2^\omega = -0.122$, $a_1^\phi = -0.738$, $a_2^\phi = 0.162$,
$a_1^{S'} = -0.0382$ and $a_2^{S'} = -0.0406$. The
QCD parameters are $\gamma=2.148$, $\Lambda^2 = 9.73$~GeV$^2$
and $Q_0^2 = 0.35$~GeV$^2$.

Apart from these changes, we adhere to the assumptions of Jaffe
\cite{bob} concerning the definition of the $\omega \phi$ mixing angle
$\epsilon$,
\beqa
|\omega\rangle &=& \cos\eps |\omega_0\rangle - \sin\eps |\phi_0\rangle
\nonumber \\
|\phi\rangle &=& \sin\eps |\omega_0\rangle + \cos\eps |\phi_0\rangle
\eeqa
as well as the parametrization of the on--mass--shell couplings of
the pure vector states ($\omega_0, \phi_0$) to the nucleons,
\beq
g_i (\phi_0 NN) \equiv g_i \sin (\eta_i) \, , \quad
g_i (\omega_0 NN) \equiv g_i \cos (\eta_i) \, \, , \eeq
with $i=1,2$ for the vector
and the tensor coupling, respectively. We also ignore SU(3)$_f$
violations in the vector meson--current couplings.\footnote{This
  assumption should eventually be relaxed in a more refined analysis.}
This universal coupling strength of each quark $q_k$ to the current
$\bar{q}_k \gamma_\mu q_k$ is denoted by $\kappa$.

To be specific, consider first the  (isoscalar) Dirac form factor $F_1 (t)$,
\beqa
& &F_1^{I=0}(t) = \Bigg[ \frac{1}{2} L^{-1}(0) + \frac{t \,\kappa g_1}
{\sqrt{6}}\Bigg( \frac{\sin(\eps + \eta_1)
\cos(\theta_0 + \eps)}{t - M_\phi^2}L^{-1}(M_\phi^2) \nonumber \\
& &- \frac{\cos(\eps+\eta_1)\sin(\theta_0+\eps)}{t - M_\omega^2}
L^{-1}(M_\omega^2)\Bigg) - \frac{t \, A_1^{I=0}}{t - M_{S'}^2}
L^{-1}(M_{S'}^2)\Bigg] L(t)
\\
& &F_1^{s}(t) = \Bigg[ -t \,\kappa g_1
\Bigg( \frac{\sin(\eps + \eta_1)
\cos\eps}{t - M_\phi^2}L^{-1}(M_\phi^2) \nonumber \\
& &- \frac{\cos(\eps+\eta_1)\sin\eps}{t - M_\omega^2}
L^{-1}(M_\omega^2)\Bigg) - \frac{t \, A_1^{s}}{t - M_{S'}^2}
L^{-1}(M_{S'}^2)\Bigg] L(t)
\label{F1}
\eeqa
with $\theta_0 = 35^\circ$ the ideal mixing angle.
The following normalization conditions and
constraints are fulfilled by construction,
\beqa
& &F_1^{I=0}(t=0)=\frac{1}{2} \, , \qquad F_1^{s}(t=0)=0 \, ,\\
& &\lim_{t \to -\infty} F_1^{I=0}(t) =0 \, , \quad
\lim_{t \to -\infty} F_1^{s}(t)=0 \,\, .
\eeqa
The constants $ A_1^{I=0}$ and $A_1^{s}$ are fixed by demanding
\beq
\lim_{t \to -\infty} L^{-1}(t) F_1^{I=0}(t) =0 \, , \quad
\lim_{t \to -\infty} L^{-1}(t) F_1^{s}(t)=0 \, \, ,
\eeq
which leads to
\beqa
A_1^{I=0}&=&L(M_{S'}^2)\Bigg[ \frac{1}{2} L^{-1}(0)
+\frac{\kappa g_1}{\sqrt{6}}\Bigg( \sin(\eps + \eta_1)
\cos(\theta_0 + \eps)L^{-1}(M_\phi^2) \nonumber \\
& &-\cos(\eps+\eta_1)\sin(\theta_0+\eps)
L^{-1}(M_\omega^2)\Bigg)\Bigg]
\\
A_1^{s}&=&L(M_{S'}^2)\Bigg[-\kappa g_1\Bigg( \sin(\eps + \eta_1)
\cos\eps \, L^{-1}(M_\phi^2) \nonumber \\
& &-\cos(\eps+\eta_1)\sin\eps \,
L^{-1}(M_\omega^2)\Bigg)\Bigg]
\eeqa
and thus the formfactors $F_1^{I=0}(t)$ and $F_1^{s}(t)$ are
determined.

In complete analogy, we parametrize the (isoscalar)
Pauli formfactor $F_2 (t)$ as
\beqa
& &F_2^{I=0}(t) = \Bigg[ \frac{\kappa g_2}{\sqrt{6}}
\Bigg( \sin(\eps + \eta_2)\cos(\theta_0 + \eps)L^{-1}(M_\phi^2)
\frac{M_\phi^2}{t - M_\phi^2}\nonumber \\
& &- \cos(\eps+\eta_2)\sin(\theta_0+\eps)L^{-1}(M_\omega^2)
\frac{M_\omega^2}{t - M_\omega^2} \Bigg)
- \frac{ M_{S'}^2 A_2^{I=0}}{t - M_{S'}^2}
L^{-1}(M_{S'}^2)\Bigg] L(t) \nonumber \\
\\
& &F_2^{s}(t) = \Bigg[ -\kappa g_2
\Bigg( \sin(\eps + \eta_2)\cos\eps\, L^{-1}(M_\phi^2)
\frac{M_\phi^2}{t - M_\phi^2}\nonumber \\
& &- \cos(\eps+\eta_2)\sin\eps\, L^{-1}(M_\omega^2)
\frac{M_\omega^2}{t - M_\omega^2} \Bigg)
- \frac{M_{S'}^2 A_2^{s}}{t - M_{S'}^2}
L^{-1}(M_{S'}^2)\Bigg] L(t)
\eeqa
subject to the constraints
\beqa
& &\lim_{t \to -\infty} F_2^{I=0}(t) =0 \, , \quad
\lim_{t \to -\infty} F_2^{s}(t)=0 \,  , \\
& &\lim_{t \to -\infty} t\, F_2^{I=0}(t) =0 \, , \quad
\lim_{t \to -\infty} t\, F_2^{s}(t)=0
\eeqa
Imposing furthermore
\beq
\lim_{t \to -\infty} L^{-1}(t)\, t\, F_2^{I=0}(t) =0 \, , \quad
\lim_{t \to -\infty} L^{-1}(t)\, t\, F_2^{s}(t)=0 \, \, ,
\eeq
leads to
\beqa
A_2^{I=0}&=&\frac{L(M_{S'}^2)}{M_{S'}^2}
\frac{\kappa g_2}{\sqrt{6}}\Bigg( \sin(\eps + \eta_2)
\cos(\theta_0 + \eps) M_\phi^2 L^{-1}(M_\phi^2) \nonumber \\
& &-\cos(\eps+\eta_2)\sin(\theta_0+\eps) M_\omega^2
L^{-1}(M_\omega^2)\Bigg)
\\
A_2^{s}&=&-\frac{L(M_{S'}^2)}{M_{S'}^2}
\kappa g_2\Bigg( \sin(\eps + \eta_2)
\cos\eps \, M_\phi^2 L^{-1}(M_\phi^2) \nonumber \\
& &-\cos(\eps+\eta_2)\sin\eps \,
M_\omega^2 L^{-1}(M_\omega^2)\Bigg)
\eeqa
and, consequently, $F_2^{I=0}(t)$ and $F_2^{s}(t)$ are given.
However, as pointed out by Musolf \cite{santorini}, the asymptotic
behaviour of the strange vector formfactos plays an important role in
such type of analyis as presented here or by Jaffe \cite{bob}. Given
the assumptions used in our analysis,
the large momentum transfer behaviour of the
strange formfactors $F^s_{1,2} (t)$ follows essentially the one of the
isoscalar electromagnetic ones, $F^{I=0}_{1,2} (t)$. Quark counting
rules suggest that the extra $\bar s s$ pair in the Fock--space
decomposition of the nucleon
wavefunction needed to describe $F^s_{1,2} (t)$ leads to a
further $t^2$--suppression as compared to the conventional isoscalar
electromagnetic formfactors.
Imposing such a constraint can lead to a significant reduction of
the strange matrix--elements at low momentum transfer
\cite{santorini}. Since our assumption about the large--$t$ fall--off
of $F^s_{1,2} (t)$ can not be excluded at present, we consider the
resulting numbers as upper limits. With this caveat in mind,
we are now in the position to analyze the strange formfactors.

\bigskip

\section{Results and discussion}
%\noindent {\bf 3.}
First, we must fix parameters. In particular, there has been some
dispute about the mixing angle $\eps$. Jaffe used the value of $\eps =
0.053 \pm 0.005$ as determined in Ref.\cite{pankaj} from $\omega ,
\phi \to 3 \pi$ decays. Since then, there have been some changes in
certain decay modes which makes
this determination to some extent uncertain. A more
elaborate treatment of symmetry breaking has been proposed by Harada
and Schechter \cite{joe}. They fit a wealth of data with a few
parameters and in that scheme $\epsilon$ is determined from
the decay mode $\phi \to
\pi^0 \gamma$, $\eps = 0.052 \ldots 0.056$. The central value used in
\cite{joe} is $\eps = 0.055$. If one ignores the effect of $\pi^0 \eta$
mixing, then $\eps$ is reduced to 0.0325. These are the benchmark values
we will use in the following.

In table~1, we show the numerical results of the fits to the nucleon
isoscalar form factors of \cite{MMD}, for the central value of $\eps$
and the very small one as discussed before. The results are stable and
within the uncertainty of Jaffe's calculation \cite{bob}.

\begin{center}

\renewcommand{\arraystretch}{1.5}

\begin{tabular}{|c||c|c|c|c|c|c|} \hline
%& & & & \\
$\eps$ & $\kappa g_1$ & $\eta_1$
& $\kappa g_2$ & $\eta_2$  & $r_2^s$ [fm$^2$] & $\mu_s$ [n.m.]
\\ \hline
0.055   & 5.36  & 0.38 & --0.93 & 0.50 & 0.21 & --0.24 \\
0.0325  & 5.48  & 0.39 & --0.95 & 0.50 & 0.23 & --0.25 \\
\hline \end{tabular}

\medskip

Table 1: Parameters and strange matrix elements extracted from
the dispersion--theoretical fit to the nucleon isoscalar
formfactors of Ref.\cite{MMD}.

\end{center}

\noindent We stress
that the fits of Ref.\cite{MMD} exhibit a much smaller variation in
the various parameters then it was the case in Ref.\cite{hoeh76} due to the
more tighter constraints. The largest uncertainty stems indeed from
the value of $\eps$. Adopting the procedure of Ref.\cite{MMD} to
estimate the uncertainties of the formfactor fits, we assign an error
of $\pm 0.03$~fm$^2$ to $r_s^2$ and of $\pm 0.03$~n.m. to $\mu_s$.
These, however, should be taken cum grano salis since we did not
consider some other sources of uncertainty like e.g. SU(3)$_f$
violation in the vector meson--current couplings.
These results are compatible with the rather uncertain determinations
of $F_{1,2}^s (0)$ from $\nu p$ elastic scattering data \cite{garvey}
(although in that paper, a negative value for $r_1^s$ is
preferred).\footnote{That analysis, however, is based on the
  assumption of one unique cut--off mass $M_A$ for all three
  axial--vector formfactors, $G^{(\alpha )} (t)$ $(\alpha = 0,3,8 )$.
  This assumption is at variance with expectations from hadron models
  which lead to a good description of the electroweak structure of the
  nucleon. For example, in Ref.\cite{bkm} it was shown that $M_A^{(0)}
  = 1.2 \, M_A^{(3)}$ and the consequences for the extraction of
  strange matrix--elements were discussed.}
Various hadron models lead to a wide range of predictions, $\mu_s
=-0.003 \ldots -0.45$~n.m. and $r_s^2 = -0.25 \ldots 0.22$~fm$^2$
\cite{vero,mudo,chn,park1,park2,weig,kim}. We remark that in the
context of the Skyrme--type models, it is often stated that the large values
of the resulting strange matrix--elements are an artefact of the
SU(3) symmetric wave functions. This deserves further study.

\vskip 2.5cm

\hskip 1in
\epsfxsize=4in
\epsfysize=3in
\epsffile{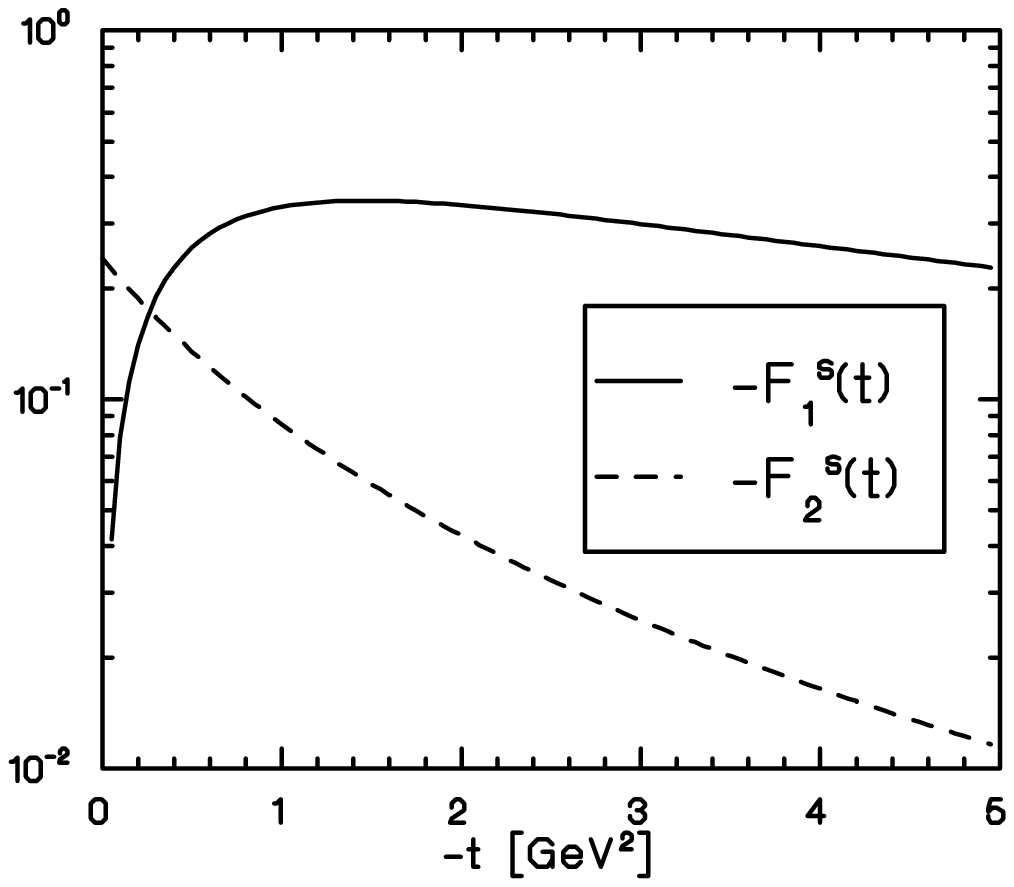}

\bigskip

\centerline{Fig.~1:\quad  The strange formfactors
  $F_{1,2}^s (t)$ for $\eps = 0.055$.}

%\smallskip}

\bigskip \bigskip

The strange formfactors $F_{1,2}^s (t)$ are shown in Fig.~1.
We note that $F_1^s (t)$ reaches its maximum in the range of momentum
transfer accessible to CEBAF. However, we stress again that
this should be considered an upper limit since a faster large--$t$
fall--off will certainly reduce the form factor \cite{santorini}.
The strange Pauli formfactor $F_2^s (t)$ can be fitted well
by a dipole,
\beq
F_2^s (t) = \mu_s \, (1 - t/2.14 \, \, {\rm GeV}^2)^{-2} \, \, ,
\eeq
i.e. with a cut-off mass of 1.46~GeV.

%\noindent
To summarize, we have updated the analysis of Ref.\cite{bob}
to deduce the  strange form factors $F_{1,2}^s (t)$ from a
dispersion--theoretical fit to the nucleons' isoscalar formfactors
\cite{MMD}. Both formfactors are negative and the strange radius and
the magnetic moment are $r_1^2 = 0.21$~fm$^2$
and $\mu_s = -0.24$~n.m., respectively. These numbers
are to be considered as upper limits due to the
large--$t$ assumptions of $F^s_{1,2} (t)$ we made.
We look forward to their experimental determinations.

\vskip 2cm

\section*{Acknowledgements}

We thank Joe Schechter and Herbert Weigel for useful correspondence.
We are particularly grateful to Mike Musolf for some clarifying comments.

\vskip 2cm

%\newpage

\end{document}